\documentclass[aps,pre,twocolumn,superscriptaddress,amsmath,amssymb]{revtex4-1}
\usepackage{graphicx}
\usepackage{float}
\usepackage{bm} 
\usepackage{verbatim}  
\usepackage{mathrsfs}
\usepackage{xcolor}

\newcommand{\ivan}[1]{\textcolor{black}{#1}}

\usepackage{dcolumn}  
\usepackage[caption=false]{subfig}
\usepackage{array}

\date{}

\begin{document}
\title{Pattern formation in active model C with anchoring: bands, aster networks, and foams}

\author{Ivan Maryshev}
\affiliation{Centre for Synthetic and Systems Biology, Institute of Cell Biology, School of Biological Sciences, The University of Edinburgh, Max Born Crescent, Edinburgh, EH9 3BF, United Kingdom}
\affiliation{SUPA, School of Physics and Astronomy, The University of Edinburgh, James Clerk Maxwell Building, Peter Guthrie Tait Road, Edinburgh, EH9 3FD, United Kingdom}

\author{Alexander Morozov}
\affiliation{SUPA, School of Physics and Astronomy, The University of Edinburgh, James Clerk Maxwell Building, Peter Guthrie Tait Road, Edinburgh, EH9 3FD, United Kingdom}

\author{Andrew B. Goryachev}
\affiliation{Centre for Synthetic and Systems Biology, Institute of Cell Biology, School of Biological Sciences, The University of Edinburgh, Max Born Crescent, Edinburgh, EH9 3BF, United Kingdom}

\author{Davide Marenduzzo}
\affiliation{SUPA, School of Physics and Astronomy, The University of Edinburgh, James Clerk Maxwell Building, Peter Guthrie Tait Road, Edinburgh, EH9 3FD, United Kingdom}

\date{\today}
\begin{abstract}
We study the dynamics of pattern formation in a minimal model for active mixtures made of microtubules and molecular motors. We monitor the evolution of the (conserved) microtubule density and of the (non-conserved) nematic order parameter, focusing on the effects of an ``anchoring'' term that provides a direct coupling between the preferred microtubule direction and their density gradient. The key control parameter is the ratio between activity and elasticity.  When elasticity dominates, the interplay between activity and anchoring leads to formation of banded structures that can undergo additional bending or rotational instabilities. When activity dominates,  the nature of anchoring instead gives rise to a range of active cellular solids, including aster-like networks, disordered foams and spindle-like patterns.  We speculate that the introduced ``active model C'' with anchoring is a minimal model to describe pattern formation in a biomimetic analogue of the microtubule cytoskeleton.
\end{abstract}

\maketitle

\section{Introduction}
A suspension of cytoskeletal filaments and motor proteins is a paradigmatic example of active matter~\cite{Marchetti2013,Needleman2017,Foster2019}. This system is active as the filaments can be moved with respect to each other by molecular motors, which consume chemical energy in the form of ATP and drive the system out of equilibrium. A typical example of such motors is given by kinesin~\cite{Kapitein2005} and its synthetic analogues that were shown to cross-link microtubules (MTs) and to push them apart.

In the last two decades, reconstituted systems containing stabilised microtubules and kinesin motors have become a standard experimental platform to investigate active phenomena~\cite{Nedelec1997,Surrey2001,Sanchez2011,Sanchez2012}. Such biomimetic systems harbour a striking variety of different patterns including extensile active bundles \cite{Sanchez2012,Decamp2015,Doostmohammadi2018}, asters \cite{Nedelec1997} or aster networks \cite{Surrey2001,Foster2015,Roostalu2018}. Notably, the same motors can organize MT filaments into different types of structures. For example, kinsein-5 can, under distinct conditions, lead to the formation of bundles, asters or networks~\cite{Roostalu2018}.  

Active behaviour of MT-motor mixtures has traditionally been explained within the active gel framework~\cite{Kruse2004}. This approach usually tacitly assumes momentum conservation and an incompressible active system. However, experimental results show that the spatial distribution of MTs can be strongly inhomogeneous, as they often form distinct dense clusters separated by almost empty spaces \cite{Henkin2014,Guillamat2016,Doostmohammadi2018}. Therefore, the MT-motor mixture is normally compressible. Additionally, in the experimentally relevant case of MT-motor mixtures bound to a substrate, the latter works as a momentum sink which quenches fluid flow, so that momentum is not conserved. 

On symmetry grounds, it is reasonable to expect a coupling between the density gradients and the orientational order parameter within a general inhomogeneous active system. Previous work has shown that extensile activity can lead to a preferential tangential alignment of MTs/active filaments at the surface~\cite{Blow2014,Maryshev2019Dry}, and that an interfacial active torque can arise when a microscopic model for a MT-motor mixture is coarse-grained. Such a torque is proportional to double gradients of the MT density $\rho$~\cite{Maryshev2019Dry}. From theories of passive liquid crystalline mixtures we also know that there is always a thermodynamic ``anchoring'', or preferential alignment, of the director field at a fluid-fluid interface~\cite{Kleman2006}. While the resulting  functional form of the interfacial active torque is different (it is bilinear in the density gradients), it has a qualitatively similar effect. 

In this work, we explore the patterns that can form in an inhomogeneous compressible active system in the presence of the liquid crystalline anchoring term. Our system is ``dry'', meaning that we do not include the solvent flow field in our equations of motion. Previously, we denoted the system in the absence of anchoring as ``active model C''~\cite{Maryshev2019Dry}, following analogy with the existing active field theory models \cite{Stenhammar2013,Tiribocchi2015}, and showed that activity leads to non-equilibrium phase separation into MT dense stripes, and, when the activity is sufficiently large, to chaotic dynamics. Here we show that the interplay between anchoring and elasticity leads to a wider range of patterns than what was found previously. Thus, for sufficiently strong normal or tangential anchoring, we find aster-like networks or active foams that are associated with multiple topological defects. Many of the patterns we find here are remarkably similar to those which self-assemble in the microtubule-motor mixtures meant as biomimetic {\it in vitro} models of microtubule cytoskeleton.

\section{Model}
Similarly to the case of passive \textit{Model C} in Hohenberg-Halperin classification~\cite{HohenbergHalperin} (HHC), we formulate our theory using the set of evolution equations for two slow variables: a conserved scalar variable $\rho(\mathbf r,t)$ describing the number density of MTs, and a tensor order parameter $Q_{ij}(\mathbf r,t)$ characterising the average nematic alignment of filaments. 

In this active \textit{Model C}, activity drives the system out of equilibrium and the underlying free energy is no longer determined. Nevertheless, the analogy with the passive limit is still of some use and in the following we will employ it where appropriate. 

The nematic $Q$-tensor that we use to describe the coarse-grained nematic order in 2D is a traceless symmetric matrix. 

We can define a nematic director field, $\mathbf n = (n_x,n_y)$, as the eigenvector corresponding to the larger (positive) eigenvalue of the $Q$-tensor. This field denotes the average orientation of individual MT filaments. For the sake of brevity we suppress the space- and time-dependence of variables in the following text. The indices $i$ and $j$ take the values of Cartesian components $x,y$.
Our equations of motion have the following general form:

\begin{align} 
\partial_t \rho =  
& 
\nabla^2
\rho^2
+ 
\chi\partial_i\partial_j \left(\rho Q_{ij}\right),
\label{rhoeq}\\
\partial_tQ_{ij} =  &  
\Big[
\left(\rho - 1\right) - \alpha Q_{kl}Q_{kl} \Big]\!Q_{ij}
+ \kappa \Delta Q _{ij} \nonumber \\
& \qquad\qquad +2\omega\left(\partial_i\rho\partial_j\rho -\frac{1}{2}\delta_{ij}\partial_k\rho \partial_k\rho \right),
\label{Qeq}
\end{align}
where $\nabla^2$ denotes the Laplacian, and $\partial_i$ and $\partial_t$ denote the spatial derivative in the $i$-th direction and the time derivative, respectively. The last term in Eq.\eqref{Qeq} is the symmetrised and traceless part of $\partial_i\rho\partial_j\rho$.

The first term in~Eq.\eqref{rhoeq} for the density evolution plays the role of a non-equilibrium chemical potential -- similar to one entering the conventional Cahn-Hilliard equation (or \textit{Model B} in HHC). It can be considered as an active contribution to the isotropic diffusion, however it has a quadratic form since typically a MM interacts with two MTs at the same time. We assume that pairwise MT interactions dominate over three-body interactions and omit cubic (or higher order) term. Thermal diffusion is not explicitly included. This is a difference with the continuum theories for self-propelled particles proposed elsewhere~\cite{Peshkov2012}. 
 
The second term in~Eq.\eqref{rhoeq} is an extensile flux of MTs coming from the motor-induced MT sliding; its strength is parametrised by $\chi$. This contribution enhances MT diffusion along the direction of nematic order and accumulates filaments in the perpendicular direction. This term was derived in~\cite{Maryshev2019Dry} by rigorously coarse-graning a microscopic model for a MM-MT mixture.
 
The terms of~Eq.\eqref{Qeq} enclosed in the square brackets provide an active analogue of the Landau terms describing concentration-induced ordering, which governs the isotropic-nematic phase transition in passive liquid crystals. The linear term defines the critical density of the isotropic-nematic transition, $\rho_{IN}$, whereas the saturating cubic term determines the equilibrium value of $Q_{ij}$ in a homogeneous nematic state. 
 
The term proportional to the Laplacian of $Q$ plays the role of elasticity in a one-elastic-constant approximation. In general, this effective elastic constant depends on the motor activity~\cite{Maryshev2019Dry}.

Finally, the last term of Eq.\eqref{Qeq} describes the non-equilibrium anchoring to the density interface in our active systems. Negative and positive values of $\omega$ correspond to parallel (tangential) and perpendicular (normal) anchoring respectively. 

Hereafter, we fix the magnitude of the saturating term in Eq.\eqref{Qeq} to $\alpha=0.05$ and independently vary the values of the free parameters $\chi,\,\kappa$, and $\omega$. Whilst the first two are always positive, the last parameter can change sign. Note that the density is normalised by the critical density of the isotropic-nematic transition. 
Additionally, the effective translational and rotational  diffusion coefficients in~Eq.\eqref{rhoeq} and \eqref{Qeq} are both equal to unity as we have scaled space and time in our equations to make them dimensionless. 


\subsection{Linear stability analysis}
Before investigating the non-linear dynamics of Eqs.~(\ref{rhoeq}-\ref{Qeq}), we consider their linear stability, as this provides a useful reference.

\begin{figure}
 \includegraphics[width=\linewidth]
 {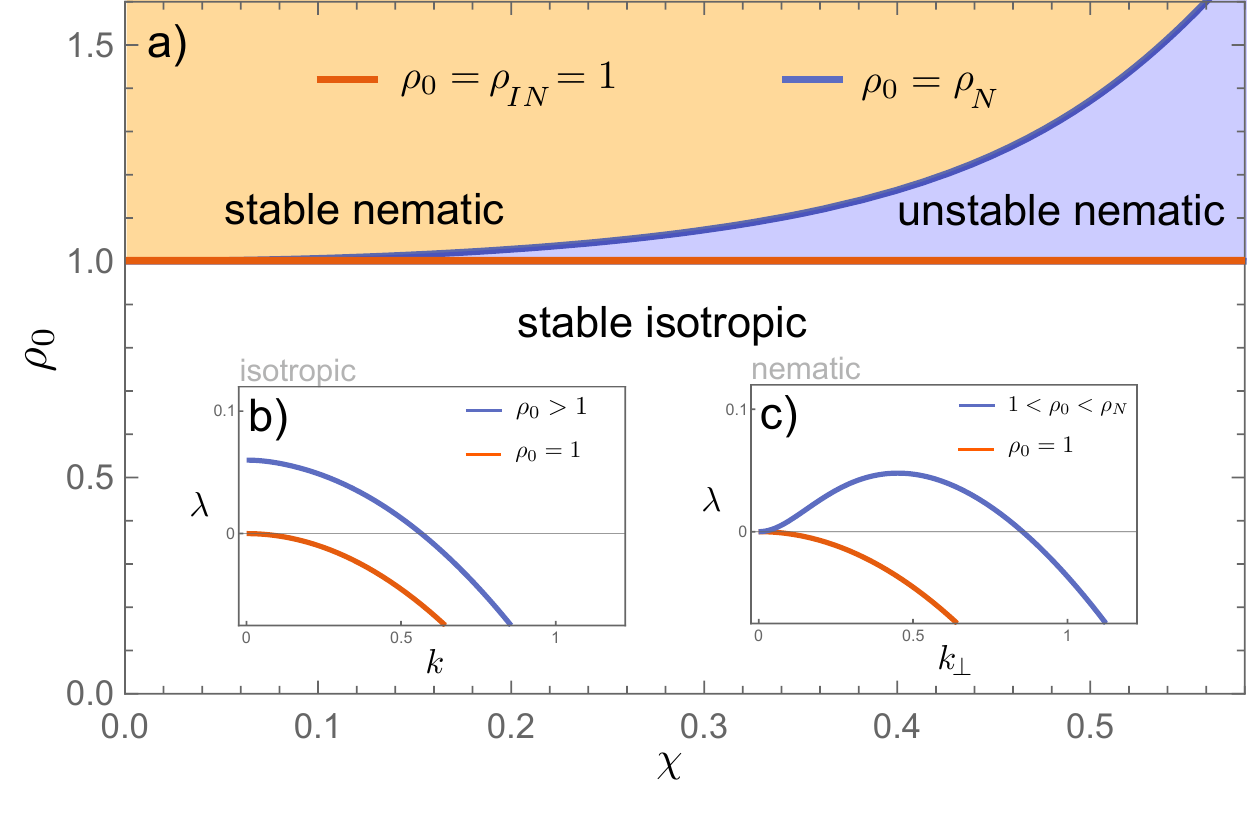}
 \caption{
 Linear stability analysis of equations~(\ref{rhoeq}-\ref{Qeq}) ($\alpha=0.05$). a) Within the white region the isotropic homogeneous state is stable. Above the red line ($\rho_{IN}=1$) the system loses stability and becomes nematic. The homogeneous nematic state is stable in the orange region for $\rho_0>\rho_N$ (above the blue line), and is unstable to phase separation within purple region. b-c) Dispersion relations for homogeneous isotropic (b) and homogeneous nematic (c) initial states ($\chi=0.5,\kappa=0.5$). The red and blue lines correspond to $\rho_0=1$ and $\rho_0=1.1$, respectively.
}
 \label{fig:LinStab}
\end{figure}

An isotropic initial state with homogeneous initial density $\rho_0$ becomes unstable to small spatial perturbations when $\rho_0>1$ (Fig.\,\ref{fig:LinStab}a). The maximal eigenvalue corresponds to growth of nematic alignment and of $Q_{ij}$. The dispersion curve depicted in Fig.\,\ref{fig:LinStab}b selects no wavelength (as the maximum is at $k=0$), so that the system spontaneously breaks the rotational symmetry and tends to form a homogeneous nematic state.

In turn, the homogeneous nematic initial state with the largest eigenvalue of the $Q$-tensor given by $Q_0=\sqrt{(\rho_0-1)/2\alpha}$ is unstable to small perturbations when $1<\rho_0<\rho_N$. In the long wavelength approximation, $\rho_N$ is a function of $\chi$ and $\alpha$.
The boundary of linear instability is defined by the following relation (Fig.~\ref{fig:LinStab}, see Appendix for details):
\begin{equation}
\chi_{cr}=\frac{4 \sqrt{2\alpha} \sqrt{\rho_N-1}\rho_N}{3\rho_N-2}.
\end{equation}
From Fig.~\ref{fig:LinStab}a it is clear that for any nonzero $\chi$, there is always a density range where the nematic phase is unstable. Within the unstable region, the dispersion relation shows that the fastest growing mode has now a well-defined wave-vector perpendicular to the orientation of filaments (Fig.\,\ref{fig:LinStab}c). At early stages of the instability, the corresponding lengthscale can be identified with a typical domain size. However, numerical simulations discussed below demonstrate that the domain size changes slowly in time and the system is prone to coarsening.

\section{Numeric results}

To explore the nonlinear dynamics of Eqs.~(\ref{rhoeq}-\ref{Qeq}) we solve them numerically. All numerical solutions are obtained by standard finite differences methods with periodic boundary conditions \ivan{\cite{AbramowitzStegun,numericalRecipes}}. The initial configuration in our simulations is either an isotropic or nematic uniform states perturbed by a small-amplitude white noise. All initial densities reported below satisfy $1<\rho_0<\rho_N$, i.e. we only consider the parameter range where the globally homogeneous nematic state is linearly unstable.

First, we consider a system without anchoring ($\omega=0$). In this case, the behaviour is determined by the microtubule density and the activity parameter $\chi$. For all $1<\rho_0<\rho_N$, filaments form dense nematically ordered bands, which are surrounded by isotropic low density regions. The spatial profiles of $\rho$ and of the positive eigenvalue of $Q_{ij}$ are approximately proportional to each other. Even in the absence of anchoring ($\omega=0$), the anisotropic nature of the active flux is sufficient to provide an effective tangential alignment of the director field at the interface. This ``active anchoring'' has been  reported previously~\cite{Maryshev2019Dry}.

The kinetic and steady state patterns, which we observe in the presence of the anchoring term, depend on the value of the elastic constant. Therefore, we now separately discuss the strong and the weak elasticity regimes that correspond to large and small values of $\kappa$, respectively. 

\subsection{Strong elasticity limit}

\begin{figure}
    \centering
    \includegraphics[width=0.999\linewidth]{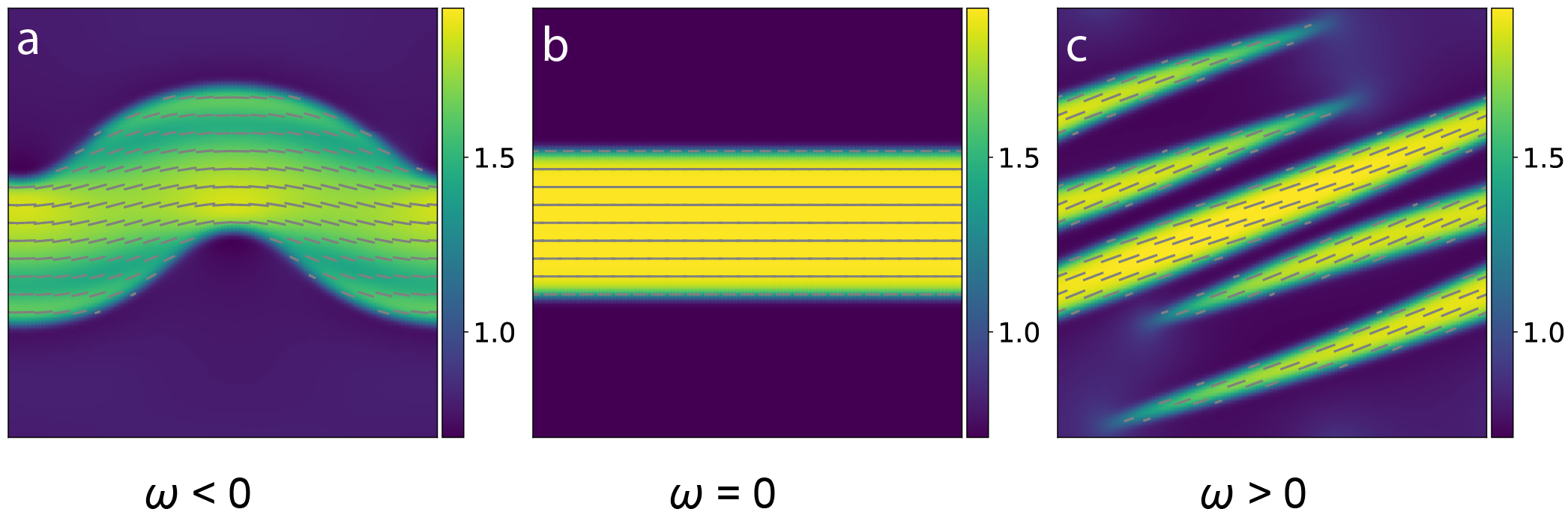}
    \caption {Numeric simulations in the strong elasticity limit:  
    (a) bending instability of the band for tangential anchoring ($\omega = -1.8$);
    (b) stable band  ($\omega = 0$),
    and (c) rotational instability of the band for normal anchoring  ($\omega = 0.45$).
    In all snapshots $\kappa=0.36,\, \rho_0=1.1,\,\chi=0.55$; colour represents $\rho$, while grey lines give the principle eigenvector  of the Q-tensor.
    } 
    \label{fig:NResults1}
\end{figure}

At large $\kappa$, the system initially forms a defect-free nematically ordered band that subsequently breaks down. Such bands become unstable via two scenarios that correspond to either tangential ($\omega<0$) or normal ($\omega>0$) anchoring (Figs.~\ref{fig:NResults1}\,a and c, S.~Movies~1 and 2). For the reasons which will become clear shortly, we will call these \emph{rotational} and \emph{bending} instabilities, respectively. 

If $\omega=0$, straight bands are always stable (see Fig.~\ref{fig:NResults1}\,b) and coarsening from a state with multiple bands is very slow in our simulations. Unlike the activity term in classic incompressible active gels~\cite{Marchetti2013}, the $\chi$ term is not, by itself (i.e., if $\omega=0$), sufficient to yield chaotic behaviour. Instead, it creates an effective tangential alignment, as discussed above. This term also controls the extent of phase separation: the stronger the anisotropic flux, the larger is the difference between minimal and maximal densities. 

When anchoring is normal ($\omega>0$), bands undergo a rotational instability. This instability arises because the normal anchoring conflicts with the tangential ordering generated by the $\chi$ term. More specifically, the imposed anchoring rotates the nematic orientation close to the interface, which in turn rotates the director in the bulk of the band due to the high elasticity that prevents gradients in $Q_{ij}$. Activity then stretches the band along the nematic director and shrinks it in the perpendicular direction, so that the band eventually breaks into several smaller ones (Fig.~\ref{fig:NResults1}\,c, S.~Movie\,2). These bands continue rotating, expanding, elongating and breaking up creating a cycle which never settles into a steady state. Bands can also merge when they meet, or occasionally disappear. In a statistically averaged steady state, there exists more than one band. This rotational instability has been previously demonstrated  for MT-MM mixtures~\cite{Maryshev2019Dry} and  for  self-propelled particles with nematic ordering~\cite{Shi2014,Cai2019}.

In the case of tangential anchoring ($\omega<0$), the active and imposed anchoring are no longer in conflict. 
Nevertheless, there is another instability that destabilises nematic bands, which we refer to as the ``bending mode''. To illustrate its mechanism, we consider a straight band that is deformed into a weakly undulating one. The sufficiently strong anchoring term favours a bend deformation, where the MT orientation within the band aligns with that at the undulating interface.  Since the anchoring term does not come from a free energy, there is no thermodynamic restoring force straightening up the band, and the undulation is instead increased by the activity $\chi$, which creates an extensile flux along the MT direction. The interplay of (non-equilibrium) anchoring and activity then acts effectively as a negative surface tension favouring an increase in the length of the interface. This means that straight bands are no longer stable and our system constantly tends to increase the curvature of interfaces, eventually resulting again in chaotic behaviour (Fig.~\ref{fig:NResults1}\,a). In line with this interpretation, we find that a straight band is linearly unstable to undulations as the strength of tangential anchoring increases (see S.~Movie\,3). We note that a similar instability was found in other models for dry active matter in~\cite{Peshkov2014,Putzig2014}, although the mechanism underlying their formation was not discussed in detail in those works.

Additionally, we find that simulations replacing the active term $\chi \partial_i\partial_j(\rho Q_{ij})$ in Eq.\eqref{rhoeq} with a term proportional to $\partial_i((\partial_j \rho)Q_{ij})$, which could be written in terms of the effective free energy, yield no patterns. This shows that the key ingredient to generate instability in Eq.\eqref{rhoeq} is the `advective' term $\chi \partial_i (\rho \partial_j(Q_{ij}))$.

In summary, the activity parameter $\chi$ drives phase separation and creates the active particle flux, whereas the anchoring is necessary to create instabilities. However, setting $\chi=0$ eliminates patterns completely as this eliminates the route to phase separation. Thus, within our model, we need both $\chi$ and $\omega$ to obtain either of the two instabilities. 

\subsection{Weak elasticity limit}

If $\kappa$ is small, then the model behaviour is different. In this case, non-equilibrium anchoring and activity can prevail over elasticity. As a result, we observe formation of topological defects, which are accompanied by density inhomogeneities.

\begin{figure*}[t!]
    \centering
    \includegraphics[width=0.7\linewidth]{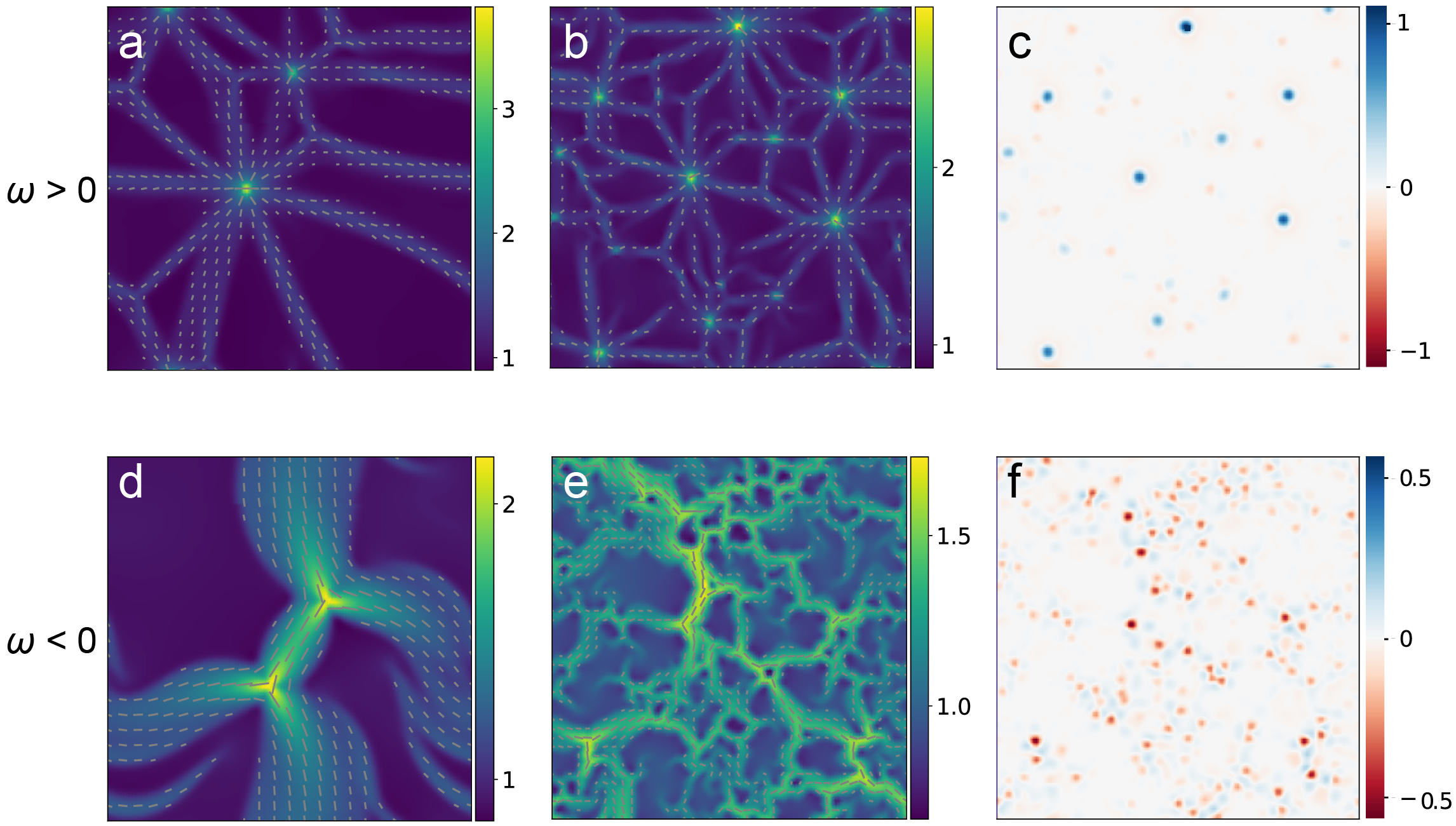}
    \caption {Numeric results for the weak elasticity regime.
    Subplots (a-c) demonstrate structures with normal anchoring:
    (a) formation of an aster with multiple discrete arms ($\omega=1.5, \, \kappa=0.05$);
    (b) active network of asters;
    (c) density of the topological charge dominated by +1 defects ($\omega=1.5, \, \kappa=0.025$).
    Subplots (d-f) represent patterns with tangential anchoring:
    (d) trefoils forming a ``doggy bone'' pair ($\omega=-1.75, \, \kappa=0.1$); 
    (e) active network of trefoils resembling ``active foam" ($\omega=-1.75, \, \kappa=0.005$);
	\ivan{(f) density of the topological charge dominated by -1/2 defects ($\omega=-1.75, \, \kappa=0.005$).
	(c) and (f) correspond to (b) and (e) respectively.
}
  In all snapshots $\rho_0=1.1,\,\chi=0.4$.
} 
    \label{fig:NResults2}
\end{figure*}

In the case of normal anchoring ($\omega>0$), the decrease of elasticity leads to the formation of topological defects coincident with peaks in MT density. In this scenario, defects have a positive topological charge of $+1$ and in the following we refer to them as \textit{asters}. High-density asters arise because the active flux pushes material towards the defect,  further increasing the density. In turn, density gradients create radial ordering of microtubules. An individual aster is associated with multiple active bundles (Fig.~\ref{fig:NResults2}\,a, S.~Movie\,4). Asters interact elastically via the MT bundles that connect them to form an active network (Fig.~\ref{fig:NResults2}\,b and c, S.~Movie\,5). For high density and small $\chi$, the network becomes much less dynamic and we observe a tendency of asters to form a hexagonal lattice.

In the case of tangential anchoring ($\omega<0$), the decrease in elasticity of undulating bands creates defects with topological charge $-1/2$. These coincide with dense trefoil-shaped patterns of $\rho$. 
As the magnitude of anchoring is increased, trefoils first form the repelling pairs having the shape of ``doggy bones'' (Fig.~\ref{fig:NResults2}\,d,  S.~Movie\,6) and then percolating foam-like networks (Fig.~\ref{fig:NResults2}\,e and f,  S.~Movie\,7). 
For high density and small $\chi$, we observe formation of an almost stable network of $-1/2$ defects. The observed networks arise as the tangential anchoring favours creation of MT rings (or hexagons) that constitute the building blocks of our active foams.

\begin{figure}[h!]
    \centering
    \includegraphics[width=0.95\linewidth]{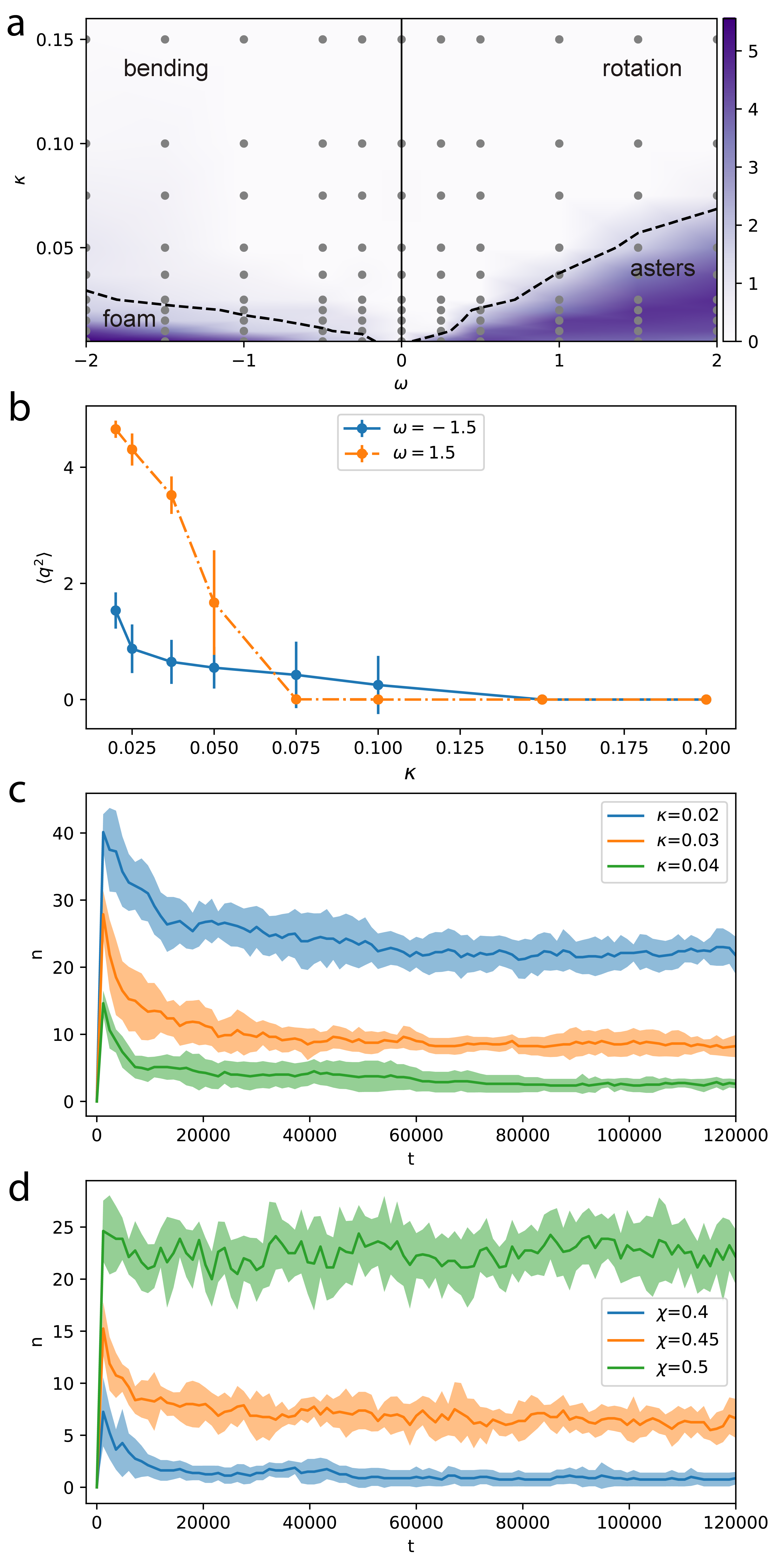}
    \caption {a-b) Time and space averaged net topological charge ($\langle q^2 \rangle$) shown as a function of $\omega$ and $\kappa$ for the fixed parameter $\chi=0.4$. 
    Plot (a) depicts a heatmap of $\langle q^2 \rangle$;  dotted line represents a nominal threshold ($\langle q^2 \rangle=1$), below which we observe formation of defects.
    Plot (b) shows the cross-sections corresponding to $\omega=\pm1.5$.
    Plots c-d) illustrate the evolution of the defect number $n$. They demonstrate the absence of coarsening, the decrease of $n$ with the growth of $\kappa$, and the increase of $n$ with the growth of $\chi$. c) Dependence of $n$ on the elastic constant $\kappa$ ($\chi=0.4,\omega=1.5$). d) Dependence of $n$ on the activity $\chi$ ($\kappa=0.05,\omega=1.5$).}
    \label{fig:Topo}
\end{figure}

To quantitatively estimate the topological charge in the system, we use the following equation for its density~\cite{Blow2014} 
\begin{equation}
    q = 
    \partial_{x} Q_{xk} \partial_{y} Q_{yk} - \partial_{x} Q_{yk} \partial_{y} Q_{xk}.
\end{equation}
Since the average of $q$ over the whole system is zero, in Fig.\ref{fig:Topo} we use $\langle q^2\rangle$ as a measure of the average topological defects number, where $\langle \cdot \rangle$ denotes spatial average over the whole system. We also average $\langle q^2\rangle$ over time, after removing the initial transient behaviour.

We observe that $\langle q^2\rangle$ increases with the decrease of the elastic constant $\kappa$ for both positive and negative $\omega$ (Fig.~4 a). This is physically intuitive, as decreasing $\kappa$ reduces the thermodynamic penalty incurred by the forming defects. This can lead to the increase in their number. 

In the case of negative $\omega$ (tangential anchoring), the transition between states with  and without defects is very smooth. In particular, for intermediate $\kappa$, we first observe formation of meta-stable trefoil defects, which later on disappear. This leads to a small value of $\langle q^2\rangle$ in this regime. However, for small $\kappa$, defects become more stable and their number quickly grows  (blue line in Fig.~4 b). 

For positive $\omega$ (normal anchoring), the transition is much sharper, and we observe no long-lived metastable structures, so that the self-assembled asters are more stable than the trefoil structures (orange line in Fig.~4 b).

To demonstrate the absence of coarsening in the patterns found at low elasticity, we explore the time evolution of the defects number. The results are shown in Fig.~3, where it can be seen that the number of defects decreases over time but remains at a finite value once a statistical steady state is reached.  In Fig.~3 we also show that for $\omega>0$ the number of positive (aster-like) defects increases with the rise in activity (Fig.~4\,c) and decreases with the increase in elasticity (Fig.~4\,d). The same is true for the negative $\omega$ (not shown).

\section{Discussion and Conclusions}

In summary, here we introduced a simple phenomenological model for pattern formation in an active filament-molecular motor mixture. Our equations of motion are minimal, but, nonetheless, they give rise to a surprising wealth of patterns, which entail nonequilibrium phase separation and nontrivial spatiotemporal dynamics. The range of patterns we find includes chaotic dynamics of active bands that continuously rotate or undulate. These patterns were found previously in several models of dry active matter~\cite{Maryshev2019Dry,Shi2014,Cai2019,Peshkov2012,Putzig2014,Putzig2016}. Importantly, we also observe previously not documented patterns with topological defects, such as networks of asters and active foams.

The key parameters of our active model C are the active extensile flux, $\chi>0$, the strength of anchoring, $\omega$, and the elastic constant $\kappa$. The sign of $\omega$ determines whether the anchoring at the interface between the high and low density regions is normal ($\omega>0$) or tangential ($\omega<0$). The anchoring created by $\omega$ is a non-equilibrium contribution: indeed a thermodynamic anchoring, which could be derived from an effective free energy, would require $\omega$ to appear in the $\rho$ equation as well. The magnitude of $\kappa$ determines the nature of the resulting patterns. For large $\kappa$, we obtain defect-free patterns with active bands, whereas for low $\kappa$ we obtain patterns with topological defects.

As our model describes extensile activity, it is applicable to mixtures of microtubules and molecular motors, such as those constituting the ``hierarchically assembled active matter'' studied experimentally in~\cite{Sanchez2012,Guillamat2016}. It is intriguing that those systems exhibit density inhomogeneities, which are reminiscent of our phase-separated microtubule bands. Bands studied in the experiments undergo a bending instability, which is qualitatively akin to the one we find with large $\kappa$ and $\omega<0$. However, as yet there is no report of observing rotational instability similar to that we find at large $\kappa$ and $\omega>0$. On the other hand, other microtubule-motor mixtures spontaneously self-organise into a network of asters {\it in vitro}~\cite{Nedelec1997,Roostalu2018}. These structures are qualitatively similar to the aster patterns we observe at low $\kappa$ and $\omega>0$. Again, these studies have not yet reported formation of active foams, such as those we find at low $\kappa$ and $\omega<0$. However, dense patterns with topological charge $-1/2$ were obtained in motility assays \cite{Huber2018}. These results indicate that not all possible combinations of $\kappa$ and $\omega$ can be realised in a real physical systems. Our results suggest that  it might be possible to switch from undulating patterns to asters in the same experimental system, if one can find a way to decrease $\kappa$ and increase $\omega$ simultaneously.

The mechanism underlying aster formation in our model is different from the one traditionally discussed in the literature\cite{YounLee2001,Maryshev2018,Aranson2006}.
In those works, asters arise as a Landau transition occurs in the polar order of filaments, while the nematic $Q$-tensor is an enslaved variable which plays a secondary role. In our case, instead, the $Q$-tensor plays the leading role. For example, when the elasticity is low, a density inhomogeneity can create an aster via the anchoring term in the equation for $Q$, and the associate active extensile flux creates filament bundles and increases the density at the aster core, stabilising the network. Although one still can define an effective polarisation $p_i$, which is determined by the instantaneous values $\partial_i\rho$ and $\partial_jQ_{ij}$, as discussed in \cite{Maryshev2019Dry}, this field is enslaved to the density and the $Q$-tensor, and plays no role in the dynamics of the system.

Importantly, asters and foams found in the active model C with anchoring presented here have not yet been reported in similar systems of equations that describe dry nematic active matter but lack the non-equilibrium anchoring contribution which we have proposed in this work. This suggests that the anchoring term might be necessary to stabilise states with non-trivial topological defects. We cannot exclude the possibility that aster networks and active foams may exist also in other versions of the active model C such as~\cite{Maryshev2019Dry}, but in a substantially smaller parameter range. Finally, we note that, while we have focused here on the case of extensile active flux ($\chi>0$ ), considering $\chi<0$ would lead to qualitatively similar patterns. Indeed, our simulations suggest that switching the signs of both $\chi$ and $\omega$ leads to the same types of patterns.

\section{Acknowledges}
AG acknowledges funding from the Biotechnology and Biological Sciences Research Council of UK (BB/P01190X, BB/P006507). DM acknowledges support from ERC CoG 648050 (THREEDCELLPHYSICS).

\section{Appendix}
Here, we discuss the linear stability analysis of the homogeneous nematic state with $\rho(\mathbf r,t)=\rho_0$, $Q_{xx}(\mathbf r,t)=-Q_{yy}(\mathbf r,t)=Q_0$, and $Q_{xy}(\mathbf r,t)=0$, where $Q_0=\sqrt{(\rho_0-1)/2\alpha}$.
The stability of this state is determined by the eigenvalues of the following matrix
\begin{align}
\left(
\begin{array}{ccc}
 -2\rho_0 k^2-\chi  \bar{k}^2 \text{Q}_0 & \chi  \bar{k}^2 \rho _0 & -2 \chi  k_x k_y \rho _0 \\
 \text{Q}_0 & -\kappa  k^2-2 (\rho_0-1) & 0 \\
 0 & 0 & -\kappa  k^2\\
\end{array}
\right),
\nonumber
\end{align}
obtained by linearising Eqs.\eqref{rhoeq} and \eqref{Qeq} around the homogeneous nematic state; perturbations are expanded in Fourier waves,  where $k$ is the corresponding wave vector, and we have also introduced $\bar{k}^2=k_x^2-k_y^2$.

We analyse the spectrum of this matrix and observe that the largest eigenvalue corresponds to a perturbation perpendicular to the direction of the nematic order ($k_x=0,k_y=k$). In this case, the system can be decomposed into two uncoupled problems, and the unstable mode is given by the eigenvalues of the following matrix
\begin{align}
\left(
\begin{array}{cc}
k^2\left( -2\rho_0+\chi Q_0\right) & \chi  k^2 \rho _0\\
Q_0 & -\kappa  k^2-2 (\rho_0-1)  \\
\end{array}
\right).
\nonumber
\end{align}
In the long-wave limit, the largest eigenvalue is real and is given by
\begin{align}
    &\lambda_{max}= k^2 \left(-2\rho_0+\chi Q_0 +\frac{\chi\rho_0 Q_0}{2(\rho_0-1)}\right)+O(k^4). \nonumber
\end{align}
Thus, the system becomes unstable when:
\begin{equation}
\chi>\frac{4\sqrt{2\alpha} \sqrt{\rho_0-1}\rho_0}{3\rho_0-2}.
\nonumber
\end{equation}

\bibliography{lit-mend}

\begin{thebibliography}{33}%
\makeatletter
\providecommand \@ifxundefined [1]{%
 \@ifx{#1\undefined}
}%
\providecommand \@ifnum [1]{%
 \ifnum #1\expandafter \@firstoftwo
 \else \expandafter \@secondoftwo
 \fi
}%
\providecommand \@ifx [1]{%
 \ifx #1\expandafter \@firstoftwo
 \else \expandafter \@secondoftwo
 \fi
}%
\providecommand \natexlab [1]{#1}%
\providecommand \enquote  [1]{``#1''}%
\providecommand \bibnamefont  [1]{#1}%
\providecommand \bibfnamefont [1]{#1}%
\providecommand \citenamefont [1]{#1}%
\providecommand \href@noop [0]{\@secondoftwo}%
\providecommand \href [0]{\begingroup \@sanitize@url \@href}%
\providecommand \@href[1]{\@@startlink{#1}\@@href}%
\providecommand \@@href[1]{\endgroup#1\@@endlink}%
\providecommand \@sanitize@url [0]{\catcode `\\12\catcode `\$12\catcode
  `\&12\catcode `\#12\catcode `\^12\catcode `\_12\catcode `\%12\relax}%
\providecommand \@@startlink[1]{}%
\providecommand \@@endlink[0]{}%
\providecommand \url  [0]{\begingroup\@sanitize@url \@url }%
\providecommand \@url [1]{\endgroup\@href {#1}{\urlprefix }}%
\providecommand \urlprefix  [0]{URL }%
\providecommand \Eprint [0]{\href }%
\providecommand \doibase [0]{http://dx.doi.org/}%
\providecommand \selectlanguage [0]{\@gobble}%
\providecommand \bibinfo  [0]{\@secondoftwo}%
\providecommand \bibfield  [0]{\@secondoftwo}%
\providecommand \translation [1]{[#1]}%
\providecommand \BibitemOpen [0]{}%
\providecommand \bibitemStop [0]{}%
\providecommand \bibitemNoStop [0]{.\EOS\space}%
\providecommand \EOS [0]{\spacefactor3000\relax}%
\providecommand \BibitemShut  [1]{\csname bibitem#1\endcsname}%
\let\auto@bib@innerbib\@empty
\bibitem [{\citenamefont {Marchetti}\ \emph {et~al.}(2013)\citenamefont
  {Marchetti}, \citenamefont {Joanny}, \citenamefont {Ramaswamy}, \citenamefont
  {Liverpool}, \citenamefont {Prost}, \citenamefont {Rao},\ and\ \citenamefont
  {Simha}}]{Marchetti2013}%
  \BibitemOpen
  \bibfield  {author} {\bibinfo {author} {\bibfnamefont {M.~C.}\ \bibnamefont
  {Marchetti}}, \bibinfo {author} {\bibfnamefont {J.-F.}\ \bibnamefont
  {Joanny}}, \bibinfo {author} {\bibfnamefont {S.}~\bibnamefont {Ramaswamy}},
  \bibinfo {author} {\bibfnamefont {T.~B.}\ \bibnamefont {Liverpool}}, \bibinfo
  {author} {\bibfnamefont {J.}~\bibnamefont {Prost}}, \bibinfo {author}
  {\bibfnamefont {M.}~\bibnamefont {Rao}}, \ and\ \bibinfo {author}
  {\bibfnamefont {R.~A.}\ \bibnamefont {Simha}},\ }\href@noop {} {\bibfield
  {journal} {\bibinfo  {journal} {Rev. Mod. Phys.}\ }\textbf {\bibinfo {volume}
  {85}},\ \bibinfo {pages} {1143} (\bibinfo {year} {2013})}\BibitemShut
  {NoStop}%
\bibitem [{\citenamefont {Needleman}\ and\ \citenamefont
  {Dogic}(2017)}]{Needleman2017}%
  \BibitemOpen
  \bibfield  {author} {\bibinfo {author} {\bibfnamefont {D.}~\bibnamefont
  {Needleman}}\ and\ \bibinfo {author} {\bibfnamefont {Z.}~\bibnamefont
  {Dogic}},\ }\href@noop {} {\bibfield  {journal} {\bibinfo  {journal} {Nat.
  Rev. Matt.}\ }\textbf {\bibinfo {volume} {2}},\ \bibinfo {pages} {17048}
  (\bibinfo {year} {2017})}\BibitemShut {NoStop}%
\bibitem [{\citenamefont {Foster}\ \emph {et~al.}(2019)\citenamefont {Foster},
  \citenamefont {F{\"u}rthauer}, \citenamefont {Shelley},\ and\ \citenamefont
  {Needleman}}]{Foster2019}%
  \BibitemOpen
  \bibfield  {author} {\bibinfo {author} {\bibfnamefont {P.~J.}\ \bibnamefont
  {Foster}}, \bibinfo {author} {\bibfnamefont {S.}~\bibnamefont
  {F{\"u}rthauer}}, \bibinfo {author} {\bibfnamefont {M.~J.}\ \bibnamefont
  {Shelley}}, \ and\ \bibinfo {author} {\bibfnamefont {D.~J.}\ \bibnamefont
  {Needleman}},\ }\href@noop {} {\bibfield  {journal} {\bibinfo  {journal}
  {Current opinion in cell biology}\ }\textbf {\bibinfo {volume} {56}},\
  \bibinfo {pages} {109} (\bibinfo {year} {2019})}\BibitemShut {NoStop}%
\bibitem [{\citenamefont {Kapitein}\ \emph {et~al.}(2005)\citenamefont
  {Kapitein}, \citenamefont {Peterman}, \citenamefont {Kwok}, \citenamefont
  {Kim}, \citenamefont {Kapoor},\ and\ \citenamefont {Schmidt}}]{Kapitein2005}%
  \BibitemOpen
  \bibfield  {author} {\bibinfo {author} {\bibfnamefont {L.~C.}\ \bibnamefont
  {Kapitein}}, \bibinfo {author} {\bibfnamefont {E.~J.}\ \bibnamefont
  {Peterman}}, \bibinfo {author} {\bibfnamefont {B.~H.}\ \bibnamefont {Kwok}},
  \bibinfo {author} {\bibfnamefont {J.~H.}\ \bibnamefont {Kim}}, \bibinfo
  {author} {\bibfnamefont {T.~M.}\ \bibnamefont {Kapoor}}, \ and\ \bibinfo
  {author} {\bibfnamefont {C.~F.}\ \bibnamefont {Schmidt}},\ }\href@noop {}
  {\bibfield  {journal} {\bibinfo  {journal} {Nature}\ }\textbf {\bibinfo
  {volume} {435}},\ \bibinfo {pages} {114} (\bibinfo {year}
  {2005})}\BibitemShut {NoStop}%
\bibitem [{\citenamefont {N{\'{e}}d{\'{e}}lec}\ \emph
  {et~al.}(1997)\citenamefont {N{\'{e}}d{\'{e}}lec}, \citenamefont {Surrey},
  \citenamefont {Maggs},\ and\ \citenamefont {Leibler}}]{Nedelec1997}%
  \BibitemOpen
  \bibfield  {author} {\bibinfo {author} {\bibfnamefont {F.~J.}\ \bibnamefont
  {N{\'{e}}d{\'{e}}lec}}, \bibinfo {author} {\bibfnamefont {T.}~\bibnamefont
  {Surrey}}, \bibinfo {author} {\bibfnamefont {A.~C.}\ \bibnamefont {Maggs}}, \
  and\ \bibinfo {author} {\bibfnamefont {S.}~\bibnamefont {Leibler}},\
  }\href@noop {} {\bibfield  {journal} {\bibinfo  {journal} {Nature}\ }\textbf
  {\bibinfo {volume} {389}},\ \bibinfo {pages} {305} (\bibinfo {year}
  {1997})}\BibitemShut {NoStop}%
\bibitem [{\citenamefont {Surrey}\ \emph {et~al.}(2001)\citenamefont {Surrey},
  \citenamefont {N{\'e}d{\'e}lec}, \citenamefont {Leibler},\ and\ \citenamefont
  {Karsenti}}]{Surrey2001}%
  \BibitemOpen
  \bibfield  {author} {\bibinfo {author} {\bibfnamefont {T.}~\bibnamefont
  {Surrey}}, \bibinfo {author} {\bibfnamefont {F.}~\bibnamefont
  {N{\'e}d{\'e}lec}}, \bibinfo {author} {\bibfnamefont {S.}~\bibnamefont
  {Leibler}}, \ and\ \bibinfo {author} {\bibfnamefont {E.}~\bibnamefont
  {Karsenti}},\ }\href@noop {} {\bibfield  {journal} {\bibinfo  {journal}
  {Science}\ }\textbf {\bibinfo {volume} {292}},\ \bibinfo {pages} {1167}
  (\bibinfo {year} {2001})}\BibitemShut {NoStop}%
\bibitem [{\citenamefont {Sanchez}\ \emph {et~al.}(2011)\citenamefont
  {Sanchez}, \citenamefont {Welch}, \citenamefont {Nicastro},\ and\
  \citenamefont {Dogic}}]{Sanchez2011}%
  \BibitemOpen
  \bibfield  {author} {\bibinfo {author} {\bibfnamefont {T.}~\bibnamefont
  {Sanchez}}, \bibinfo {author} {\bibfnamefont {D.}~\bibnamefont {Welch}},
  \bibinfo {author} {\bibfnamefont {D.}~\bibnamefont {Nicastro}}, \ and\
  \bibinfo {author} {\bibfnamefont {Z.}~\bibnamefont {Dogic}},\ }\href@noop {}
  {\bibfield  {journal} {\bibinfo  {journal} {Science}\ }\textbf {\bibinfo
  {volume} {333}},\ \bibinfo {pages} {456 } (\bibinfo {year}
  {2011})}\BibitemShut {NoStop}%
\bibitem [{\citenamefont {Sanchez}\ \emph {et~al.}(2012)\citenamefont
  {Sanchez}, \citenamefont {Chen}, \citenamefont {Decamp}, \citenamefont
  {Heymann},\ and\ \citenamefont {Dogic}}]{Sanchez2012}%
  \BibitemOpen
  \bibfield  {author} {\bibinfo {author} {\bibfnamefont {T.}~\bibnamefont
  {Sanchez}}, \bibinfo {author} {\bibfnamefont {D.~T.~N.}\ \bibnamefont
  {Chen}}, \bibinfo {author} {\bibfnamefont {S.~J.}\ \bibnamefont {Decamp}},
  \bibinfo {author} {\bibfnamefont {M.}~\bibnamefont {Heymann}}, \ and\
  \bibinfo {author} {\bibfnamefont {Z.}~\bibnamefont {Dogic}},\ }\href@noop {}
  {\bibfield  {journal} {\bibinfo  {journal} {Nature}\ }\textbf {\bibinfo
  {volume} {491}},\ \bibinfo {pages} {1} (\bibinfo {year} {2012})}\BibitemShut
  {NoStop}%
\bibitem [{\citenamefont {DeCamp}\ \emph {et~al.}(2015)\citenamefont {DeCamp},
  \citenamefont {Redner}, \citenamefont {Baskaran}, \citenamefont {Hagan},\
  and\ \citenamefont {Dogic}}]{Decamp2015}%
  \BibitemOpen
  \bibfield  {author} {\bibinfo {author} {\bibfnamefont {S.~J.}\ \bibnamefont
  {DeCamp}}, \bibinfo {author} {\bibfnamefont {G.~S.}\ \bibnamefont {Redner}},
  \bibinfo {author} {\bibfnamefont {A.}~\bibnamefont {Baskaran}}, \bibinfo
  {author} {\bibfnamefont {M.~F.}\ \bibnamefont {Hagan}}, \ and\ \bibinfo
  {author} {\bibfnamefont {Z.}~\bibnamefont {Dogic}},\ }\href@noop {}
  {\bibfield  {journal} {\bibinfo  {journal} {Nat. Mater.}\ }\textbf {\bibinfo
  {volume} {14}},\ \bibinfo {pages} {1110} (\bibinfo {year}
  {2015})}\BibitemShut {NoStop}%
\bibitem [{\citenamefont {Doostmohammadi}\ \emph {et~al.}(2018)\citenamefont
  {Doostmohammadi}, \citenamefont {Ign{\'e}s-Mullol}, \citenamefont {Yeomans},\
  and\ \citenamefont {Sagu{\'e}s}}]{Doostmohammadi2018}%
  \BibitemOpen
  \bibfield  {author} {\bibinfo {author} {\bibfnamefont {A.}~\bibnamefont
  {Doostmohammadi}}, \bibinfo {author} {\bibfnamefont {J.}~\bibnamefont
  {Ign{\'e}s-Mullol}}, \bibinfo {author} {\bibfnamefont {J.~M.}\ \bibnamefont
  {Yeomans}}, \ and\ \bibinfo {author} {\bibfnamefont {F.}~\bibnamefont
  {Sagu{\'e}s}},\ }\href@noop {} {\bibfield  {journal} {\bibinfo  {journal}
  {Nat. Commun.}\ }\textbf {\bibinfo {volume} {9}},\ \bibinfo {pages} {3246}
  (\bibinfo {year} {2018})}\BibitemShut {NoStop}%
\bibitem [{\citenamefont {Foster}\ \emph {et~al.}(2015)\citenamefont {Foster},
  \citenamefont {F{\"{u}}rthauer}, \citenamefont {Shelley},\ and\ \citenamefont
  {Needleman}}]{Foster2015}%
  \BibitemOpen
  \bibfield  {author} {\bibinfo {author} {\bibfnamefont {P.~J.}\ \bibnamefont
  {Foster}}, \bibinfo {author} {\bibfnamefont {S.}~\bibnamefont
  {F{\"{u}}rthauer}}, \bibinfo {author} {\bibfnamefont {M.~J.}\ \bibnamefont
  {Shelley}}, \ and\ \bibinfo {author} {\bibfnamefont {D.~J.}\ \bibnamefont
  {Needleman}},\ }\href {\doibase 10.7554/eLife.10837} {\bibfield  {journal}
  {\bibinfo  {journal} {Elife}\ }\textbf {\bibinfo {volume} {4}},\ \bibinfo
  {pages} {e10837} (\bibinfo {year} {2015})}\BibitemShut {NoStop}%
\bibitem [{\citenamefont {Roostalu}\ \emph {et~al.}(2018)\citenamefont
  {Roostalu}, \citenamefont {Rickman}, \citenamefont {Thomas}, \citenamefont
  {N{\'e}d{\'e}lec},\ and\ \citenamefont {Surrey}}]{Roostalu2018}%
  \BibitemOpen
  \bibfield  {author} {\bibinfo {author} {\bibfnamefont {J.}~\bibnamefont
  {Roostalu}}, \bibinfo {author} {\bibfnamefont {J.}~\bibnamefont {Rickman}},
  \bibinfo {author} {\bibfnamefont {C.}~\bibnamefont {Thomas}}, \bibinfo
  {author} {\bibfnamefont {F.}~\bibnamefont {N{\'e}d{\'e}lec}}, \ and\ \bibinfo
  {author} {\bibfnamefont {T.}~\bibnamefont {Surrey}},\ }\href@noop {}
  {\bibfield  {journal} {\bibinfo  {journal} {Cell}\ }\textbf {\bibinfo
  {volume} {175}},\ \bibinfo {pages} {796} (\bibinfo {year}
  {2018})}\BibitemShut {NoStop}%
\bibitem [{\citenamefont {Kruse}\ \emph {et~al.}(2004)\citenamefont {Kruse},
  \citenamefont {Joanny}, \citenamefont {J{\"{u}}licher}, \citenamefont
  {Prost},\ and\ \citenamefont {Sekimoto}}]{Kruse2004}%
  \BibitemOpen
  \bibfield  {author} {\bibinfo {author} {\bibfnamefont {K.}~\bibnamefont
  {Kruse}}, \bibinfo {author} {\bibfnamefont {J.~F.}\ \bibnamefont {Joanny}},
  \bibinfo {author} {\bibfnamefont {F.}~\bibnamefont {J{\"{u}}licher}},
  \bibinfo {author} {\bibfnamefont {J.}~\bibnamefont {Prost}}, \ and\ \bibinfo
  {author} {\bibfnamefont {K.}~\bibnamefont {Sekimoto}},\ }\href {\doibase
  10.1103/PhysRevLett.93.099902} {\bibfield  {journal} {\bibinfo  {journal}
  {Phys. Rev. Lett.}\ }\textbf {\bibinfo {volume} {92}},\ \bibinfo {pages}
  {078101} (\bibinfo {year} {2004})}\BibitemShut {NoStop}%
\bibitem [{\citenamefont {Henkin}\ \emph {et~al.}(2014)\citenamefont {Henkin},
  \citenamefont {DeCamp}, \citenamefont {Chen}, \citenamefont {Sanchez},\ and\
  \citenamefont {Dogic}}]{Henkin2014}%
  \BibitemOpen
  \bibfield  {author} {\bibinfo {author} {\bibfnamefont {G.}~\bibnamefont
  {Henkin}}, \bibinfo {author} {\bibfnamefont {S.~J.}\ \bibnamefont {DeCamp}},
  \bibinfo {author} {\bibfnamefont {D.~T.~N.}\ \bibnamefont {Chen}}, \bibinfo
  {author} {\bibfnamefont {T.}~\bibnamefont {Sanchez}}, \ and\ \bibinfo
  {author} {\bibfnamefont {Z.}~\bibnamefont {Dogic}},\ }\href@noop {}
  {\bibfield  {journal} {\bibinfo  {journal} {Phil. Trans. R. Soc. A}\ }\textbf
  {\bibinfo {volume} {372}},\ \bibinfo {pages} {20140142} (\bibinfo {year}
  {2014})}\BibitemShut {NoStop}%
\bibitem [{\citenamefont {Guillamat}\ \emph {et~al.}(2016)\citenamefont
  {Guillamat}, \citenamefont {Ign{\'{e}}s-Mullol},\ and\ \citenamefont
  {Sagu{\'{e}}s}}]{Guillamat2016}%
  \BibitemOpen
  \bibfield  {author} {\bibinfo {author} {\bibfnamefont {P.}~\bibnamefont
  {Guillamat}}, \bibinfo {author} {\bibfnamefont {J.}~\bibnamefont
  {Ign{\'{e}}s-Mullol}}, \ and\ \bibinfo {author} {\bibfnamefont
  {F.}~\bibnamefont {Sagu{\'{e}}s}},\ }\href@noop {} {\bibfield  {journal}
  {\bibinfo  {journal} {Proc. Natl. Acad. Sci.}\ }\textbf {\bibinfo {volume}
  {113}},\ \bibinfo {pages} {5498} (\bibinfo {year} {2016})}\BibitemShut
  {NoStop}%
\bibitem [{\citenamefont {Blow}\ \emph {et~al.}(2014)\citenamefont {Blow},
  \citenamefont {Thampi},\ and\ \citenamefont {Yeomans}}]{Blow2014}%
  \BibitemOpen
  \bibfield  {author} {\bibinfo {author} {\bibfnamefont {M.~L.}\ \bibnamefont
  {Blow}}, \bibinfo {author} {\bibfnamefont {S.~P.}\ \bibnamefont {Thampi}}, \
  and\ \bibinfo {author} {\bibfnamefont {J.~M.}\ \bibnamefont {Yeomans}},\
  }\href@noop {} {\bibfield  {journal} {\bibinfo  {journal} {Phys. Rev. Lett.}\
  }\textbf {\bibinfo {volume} {113}},\ \bibinfo {pages} {248303} (\bibinfo
  {year} {2014})}\BibitemShut {NoStop}%
\bibitem [{\citenamefont {Maryshev}\ \emph {et~al.}(2019)\citenamefont
  {Maryshev}, \citenamefont {Goryachev}, \citenamefont {Marenduzzo},\ and\
  \citenamefont {Morozov}}]{Maryshev2019Dry}%
  \BibitemOpen
  \bibfield  {author} {\bibinfo {author} {\bibfnamefont {I.}~\bibnamefont
  {Maryshev}}, \bibinfo {author} {\bibfnamefont {A.~B.}\ \bibnamefont
  {Goryachev}}, \bibinfo {author} {\bibfnamefont {D.}~\bibnamefont
  {Marenduzzo}}, \ and\ \bibinfo {author} {\bibfnamefont {A.}~\bibnamefont
  {Morozov}},\ }\href@noop {} {\bibfield  {journal} {\bibinfo  {journal} {Soft
  Matter}\ }\textbf {\bibinfo {volume} {15}},\ \bibinfo {pages} {6038}
  (\bibinfo {year} {2019})}\BibitemShut {NoStop}%
\bibitem [{\citenamefont {Kleman}\ and\ \citenamefont
  {Lavrentovich}(2006)}]{Kleman2006}%
  \BibitemOpen
  \bibfield  {author} {\bibinfo {author} {\bibfnamefont {M.}~\bibnamefont
  {Kleman}}\ and\ \bibinfo {author} {\bibfnamefont {O.~D.}\ \bibnamefont
  {Lavrentovich}},\ }\href@noop {} {\bibfield  {journal} {\bibinfo  {journal}
  {Philos. Mag.}\ }\textbf {\bibinfo {volume} {86}},\ \bibinfo {pages} {4117}
  (\bibinfo {year} {2006})}\BibitemShut {NoStop}%
\bibitem [{\citenamefont {Stenhammar}\ \emph {et~al.}(2013)\citenamefont
  {Stenhammar}, \citenamefont {Tiribocchi}, \citenamefont {Allen},
  \citenamefont {Marenduzzo},\ and\ \citenamefont {Cates}}]{Stenhammar2013}%
  \BibitemOpen
  \bibfield  {author} {\bibinfo {author} {\bibfnamefont {J.}~\bibnamefont
  {Stenhammar}}, \bibinfo {author} {\bibfnamefont {A.}~\bibnamefont
  {Tiribocchi}}, \bibinfo {author} {\bibfnamefont {R.~J.}\ \bibnamefont
  {Allen}}, \bibinfo {author} {\bibfnamefont {D.}~\bibnamefont {Marenduzzo}}, \
  and\ \bibinfo {author} {\bibfnamefont {M.~E.}\ \bibnamefont {Cates}},\
  }\href@noop {} {\bibfield  {journal} {\bibinfo  {journal} {Phys. Rev. Lett.}\
  }\textbf {\bibinfo {volume} {111}},\ \bibinfo {pages} {145702} (\bibinfo
  {year} {2013})}\BibitemShut {NoStop}%
\bibitem [{\citenamefont {Tiribocchi}\ \emph {et~al.}(2015)\citenamefont
  {Tiribocchi}, \citenamefont {Wittkowski}, \citenamefont {Marenduzzo},\ and\
  \citenamefont {Cates}}]{Tiribocchi2015}%
  \BibitemOpen
  \bibfield  {author} {\bibinfo {author} {\bibfnamefont {A.}~\bibnamefont
  {Tiribocchi}}, \bibinfo {author} {\bibfnamefont {R.}~\bibnamefont
  {Wittkowski}}, \bibinfo {author} {\bibfnamefont {D.}~\bibnamefont
  {Marenduzzo}}, \ and\ \bibinfo {author} {\bibfnamefont {M.~E.}\ \bibnamefont
  {Cates}},\ }\href@noop {} {\bibfield  {journal} {\bibinfo  {journal} {Phys.
  Rev. Lett.}\ }\textbf {\bibinfo {volume} {115}},\ \bibinfo {pages} {188302}
  (\bibinfo {year} {2015})}\BibitemShut {NoStop}%
\bibitem [{\citenamefont {Hohenberg}\ and\ \citenamefont
  {Halperin}(1977)}]{HohenbergHalperin}%
  \BibitemOpen
  \bibfield  {author} {\bibinfo {author} {\bibfnamefont {P.~C.}\ \bibnamefont
  {Hohenberg}}\ and\ \bibinfo {author} {\bibfnamefont {B.~I.}\ \bibnamefont
  {Halperin}},\ }\href@noop {} {\bibfield  {journal} {\bibinfo  {journal} {Rev.
  Mod. Phys.}\ }\textbf {\bibinfo {volume} {49}},\ \bibinfo {pages} {435}
  (\bibinfo {year} {1977})}\BibitemShut {NoStop}%
\bibitem [{\citenamefont {Peshkov}\ \emph {et~al.}(2012)\citenamefont
  {Peshkov}, \citenamefont {Aranson}, \citenamefont {Bertin}, \citenamefont
  {Chat{\'{e}}},\ and\ \citenamefont {Ginelli}}]{Peshkov2012}%
  \BibitemOpen
  \bibfield  {author} {\bibinfo {author} {\bibfnamefont {A.}~\bibnamefont
  {Peshkov}}, \bibinfo {author} {\bibfnamefont {I.~S.}\ \bibnamefont
  {Aranson}}, \bibinfo {author} {\bibfnamefont {E.}~\bibnamefont {Bertin}},
  \bibinfo {author} {\bibfnamefont {H.}~\bibnamefont {Chat{\'{e}}}}, \ and\
  \bibinfo {author} {\bibfnamefont {F.}~\bibnamefont {Ginelli}},\ }\href@noop
  {} {\bibfield  {journal} {\bibinfo  {journal} {Phys. Rev. Lett.}\ }\textbf
  {\bibinfo {volume} {109}},\ \bibinfo {pages} {268701} (\bibinfo {year}
  {2012})}\BibitemShut {NoStop}%
\bibitem [{\citenamefont {Abramowitz}\ and\ \citenamefont
  {Stegun}(1964)}]{AbramowitzStegun}%
  \BibitemOpen
  \bibfield  {author} {\bibinfo {author} {\bibfnamefont {M.}~\bibnamefont
  {Abramowitz}}\ and\ \bibinfo {author} {\bibfnamefont {I.~A.}\ \bibnamefont
  {Stegun}},\ }\href@noop {} {\emph {\bibinfo {title} {Handbook of Mathematical
  Functions with Formulas, Graphs, and Mathematical Tables}}}\ (\bibinfo
  {publisher} {Dover},\ \bibinfo {address} {New York},\ \bibinfo {year}
  {1964})\BibitemShut {NoStop}%
\bibitem [{\citenamefont {Press}\ \emph {et~al.}(2007)\citenamefont {Press},
  \citenamefont {Teukolsky}, \citenamefont {Vetterling},\ and\ \citenamefont
  {Flannery}}]{numericalRecipes}%
  \BibitemOpen
  \bibfield  {author} {\bibinfo {author} {\bibfnamefont {W.~H.}\ \bibnamefont
  {Press}}, \bibinfo {author} {\bibfnamefont {S.~A.}\ \bibnamefont
  {Teukolsky}}, \bibinfo {author} {\bibfnamefont {W.~T.}\ \bibnamefont
  {Vetterling}}, \ and\ \bibinfo {author} {\bibfnamefont {B.~P.}\ \bibnamefont
  {Flannery}},\ }\href@noop {} {\emph {\bibinfo {title} {Numerical Recipes 3rd
  Edition: The Art of Scientific Computing}}},\ \bibinfo {edition} {3rd}\ ed.\
  (\bibinfo  {publisher} {Cambridge University Press},\ \bibinfo {address} {New
  York, NY, USA},\ \bibinfo {year} {2007})\BibitemShut {NoStop}%
\bibitem [{\citenamefont {Shi}\ \emph {et~al.}(2014)\citenamefont {Shi},
  \citenamefont {Chat{\'{e}}},\ and\ \citenamefont {Ma}}]{Shi2014}%
  \BibitemOpen
  \bibfield  {author} {\bibinfo {author} {\bibfnamefont {X.-Q.~Q.}\
  \bibnamefont {Shi}}, \bibinfo {author} {\bibfnamefont {H.}~\bibnamefont
  {Chat{\'{e}}}}, \ and\ \bibinfo {author} {\bibfnamefont {Y.-Q.~Q.}\
  \bibnamefont {Ma}},\ }\href {\doibase 10.1088/1367-2630/16/3/035003}
  {\bibfield  {journal} {\bibinfo  {journal} {New J. Phys.}\ }\textbf {\bibinfo
  {volume} {16}},\ \bibinfo {pages} {35003} (\bibinfo {year}
  {2014})}\BibitemShut {NoStop}%
\bibitem [{\citenamefont {Cai}\ \emph {et~al.}(2019)\citenamefont {Cai},
  \citenamefont {Chat\'e}, \citenamefont {Ma},\ and\ \citenamefont
  {Shi}}]{Cai2019}%
  \BibitemOpen
  \bibfield  {author} {\bibinfo {author} {\bibfnamefont {L.-B.}\ \bibnamefont
  {Cai}}, \bibinfo {author} {\bibfnamefont {H.}~\bibnamefont {Chat\'e}},
  \bibinfo {author} {\bibfnamefont {Y.-Q.}\ \bibnamefont {Ma}}, \ and\ \bibinfo
  {author} {\bibfnamefont {X.-Q.}\ \bibnamefont {Shi}},\ }\href@noop {}
  {\bibfield  {journal} {\bibinfo  {journal} {Phys. Rev. E}\ }\textbf {\bibinfo
  {volume} {99}},\ \bibinfo {pages} {010601} (\bibinfo {year}
  {2019})}\BibitemShut {NoStop}%
\bibitem [{\citenamefont {Peshkov}\ \emph {et~al.}(2014)\citenamefont
  {Peshkov}, \citenamefont {Bertin}, \citenamefont {Ginelli},\ and\
  \citenamefont {Chat{\'{e}}}}]{Peshkov2014}%
  \BibitemOpen
  \bibfield  {author} {\bibinfo {author} {\bibfnamefont {A.}~\bibnamefont
  {Peshkov}}, \bibinfo {author} {\bibfnamefont {E.}~\bibnamefont {Bertin}},
  \bibinfo {author} {\bibfnamefont {F.}~\bibnamefont {Ginelli}}, \ and\
  \bibinfo {author} {\bibfnamefont {H.}~\bibnamefont {Chat{\'{e}}}},\
  }\href@noop {} {\bibfield  {journal} {\bibinfo  {journal} {Eur. Phys. J.
  Spec. Top.}\ }\textbf {\bibinfo {volume} {223}},\ \bibinfo {pages} {1315}
  (\bibinfo {year} {2014})}\BibitemShut {NoStop}%
\bibitem [{\citenamefont {Putzig}\ and\ \citenamefont
  {Baskaran}(2014)}]{Putzig2014}%
  \BibitemOpen
  \bibfield  {author} {\bibinfo {author} {\bibfnamefont {E.}~\bibnamefont
  {Putzig}}\ and\ \bibinfo {author} {\bibfnamefont {A.}~\bibnamefont
  {Baskaran}},\ }\href@noop {} {\bibfield  {journal} {\bibinfo  {journal}
  {Phys. Rev. E}\ }\textbf {\bibinfo {volume} {90}},\ \bibinfo {pages} {042304}
  (\bibinfo {year} {2014})}\BibitemShut {NoStop}%
\bibitem [{\citenamefont {Putzig}\ \emph {et~al.}(2016)\citenamefont {Putzig},
  \citenamefont {Redner}, \citenamefont {Baskaran},\ and\ \citenamefont
  {Baskaran}}]{Putzig2016}%
  \BibitemOpen
  \bibfield  {author} {\bibinfo {author} {\bibfnamefont {E.}~\bibnamefont
  {Putzig}}, \bibinfo {author} {\bibfnamefont {G.~S.}\ \bibnamefont {Redner}},
  \bibinfo {author} {\bibfnamefont {A.}~\bibnamefont {Baskaran}}, \ and\
  \bibinfo {author} {\bibfnamefont {A.}~\bibnamefont {Baskaran}},\ }\href@noop
  {} {\bibfield  {journal} {\bibinfo  {journal} {Soft Matter}\ }\textbf
  {\bibinfo {volume} {12}},\ \bibinfo {pages} {3854} (\bibinfo {year}
  {2016})}\BibitemShut {NoStop}%
\bibitem [{\citenamefont {Huber}\ \emph {et~al.}(2018)\citenamefont {Huber},
  \citenamefont {Suzuki}, \citenamefont {Kr{\"u}ger}, \citenamefont {Frey},\
  and\ \citenamefont {Bausch}}]{Huber2018}%
  \BibitemOpen
  \bibfield  {author} {\bibinfo {author} {\bibfnamefont {L.}~\bibnamefont
  {Huber}}, \bibinfo {author} {\bibfnamefont {R.}~\bibnamefont {Suzuki}},
  \bibinfo {author} {\bibfnamefont {T.}~\bibnamefont {Kr{\"u}ger}}, \bibinfo
  {author} {\bibfnamefont {E.}~\bibnamefont {Frey}}, \ and\ \bibinfo {author}
  {\bibfnamefont {A.}~\bibnamefont {Bausch}},\ }\href@noop {} {\bibfield
  {journal} {\bibinfo  {journal} {Science}\ }\textbf {\bibinfo {volume}
  {361}},\ \bibinfo {pages} {255} (\bibinfo {year} {2018})}\BibitemShut
  {NoStop}%
\bibitem [{\citenamefont {Lee}\ and\ \citenamefont
  {Kardar}(2001)}]{YounLee2001}%
  \BibitemOpen
  \bibfield  {author} {\bibinfo {author} {\bibfnamefont {H.~Y.}\ \bibnamefont
  {Lee}}\ and\ \bibinfo {author} {\bibfnamefont {M.}~\bibnamefont {Kardar}},\
  }\href@noop {} {\bibfield  {journal} {\bibinfo  {journal} {Phys. Rev. E}\
  }\textbf {\bibinfo {volume} {64}},\ \bibinfo {pages} {056113} (\bibinfo
  {year} {2001})}\BibitemShut {NoStop}%
\bibitem [{\citenamefont {Maryshev}\ \emph {et~al.}(2018)\citenamefont
  {Maryshev}, \citenamefont {Marenduzzo}, \citenamefont {Goryachev},\ and\
  \citenamefont {Morozov}}]{Maryshev2018}%
  \BibitemOpen
  \bibfield  {author} {\bibinfo {author} {\bibfnamefont {I.}~\bibnamefont
  {Maryshev}}, \bibinfo {author} {\bibfnamefont {D.}~\bibnamefont
  {Marenduzzo}}, \bibinfo {author} {\bibfnamefont {A.~B.}\ \bibnamefont
  {Goryachev}}, \ and\ \bibinfo {author} {\bibfnamefont {A.}~\bibnamefont
  {Morozov}},\ }\href {\doibase 10.1103/PhysRevE.97.022412} {\bibfield
  {journal} {\bibinfo  {journal} {Phys. Rev. E}\ }\textbf {\bibinfo {volume}
  {97}},\ \bibinfo {pages} {22412} (\bibinfo {year} {2018})}\BibitemShut
  {NoStop}%
\bibitem [{\citenamefont {Aranson}\ and\ \citenamefont
  {Tsimring}(2006)}]{Aranson2006}%
  \BibitemOpen
  \bibfield  {author} {\bibinfo {author} {\bibfnamefont {I.~S.}\ \bibnamefont
  {Aranson}}\ and\ \bibinfo {author} {\bibfnamefont {L.~S.}\ \bibnamefont
  {Tsimring}},\ }\href {\doibase 10.1103/PhysRevE.74.031915} {\bibfield
  {journal} {\bibinfo  {journal} {Phys. Rev. E}\ }\textbf {\bibinfo {volume}
  {74}},\ \bibinfo {pages} {31915} (\bibinfo {year} {2006})}\BibitemShut
  {NoStop}%
\end{thebibliography}%


%
\end{document}